# Fast spectroscopic mapping of two-dimensional quantum materials

**Authors:** Berk Zengin[1], Jens Oppliger[1], Danyang Liu[1], Lorena Niggli[1], Tohru Kurosawa[2], and Fabian D. Natterer[1]*

**Affiliations:**

[1]Department of Physics, University of Zurich, Winterthurerstrasse 190, 8057 Zurich, Switzerland

[2]Department of Physics, Hokkaido University, Sapporo 060-0810, Japan

*Correspondence to: Fabian D. Natterer (fabian.natterer@uzh.ch)

**Abstract:**

Spectroscopic mapping refers to the massive recording of spectra whilst varying an additional degree of freedom, such as: magnetic field, location, temperature, or charge carrier concentration. As this involves two serial tasks, spectroscopic mapping can become excruciatingly slow. We demonstrate exponentially faster mapping through our combination of sparse sampling and parallel spectroscopy. We exemplify our concept using quasiparticle interference imaging of Au(111) and $Bi_2Sr_2CaCu_2O_{8+\delta}$ (Bi2212), as two well-known model systems. Our method is accessible, straightforward to implement with existing scanning tunneling microscopes, and can be easily extended to enhance gate or field-mapping spectroscopy. In view of a possible $10^4$-fold speed advantage, it is setting the stage to fundamentally promote the discovery of novel quantum materials.

**One Sentence Summary:** Versatile method to rapidly discover the properties of quantum materials with a scanning tunneling microscope.

**Main Text:**

The late Freeman Dyson marveled at the luxury of our times in which scientific progress is strong both in idea and tool driven research, one occasionally outdoing the other(*1*). The field of quantum materials(*2*) is oddly related to these observations because it has been heavily pushed by theoretical advances to which their experimental counterpart is still playing catch-up. This is partially owed to the rapid development of numerical methods that allow for an inspection of a vast parameter space in the hunt for novel materials but also due to the hard realities in their experimental discovery that deal with real constraints, such as: degradation, sample size, environment, and the lack of proper characterization tools. The active feedback from experimental observations would positively guide the theoretical exploration of new candidate materials and it exposes the urgent need for suitable characterization tools to enable and keep up with future developments.

We examine the necessity for novel concepts in the field of spectroscopic mapping with the scanning tunneling microscope (STM) as a tool in the discovery of quantum materials. From the beginning of its conception, the STM has been both, notorious for its slow measurement pace that is intrinsic to the concept, and also revered for continually evolving into the tool that provides essential insights into the workings of the microscopic world(*3–6*). To expose an Achilles heel of STM investigations, we exemplify here one prominent measurement mode that involves the combination of topography and spectroscopy. Note that other STM modes, such as gate mapping spectroscopy(*6*) or field-dependent spectroscopy(*7*), experience similar issues that can be equally treated within the context laid out in this contribution. What unites these modes is their combination of at least two serial tasks that involve recording the spectroscopy for every increment of an external parameter, which can be: topography, temperature, magnetic field, doping, or any combination thereof. The enchaining of two serial tasks is what renders spectroscopic mapping excruciatingly slow and oftentimes imposes compromises that directly impact data quality or spectroscopic resolution.

The total mapping time depends on the product of time per spectrum and the number of recorded spectra, i.e., the number of increments of the external parameter. The trivial conclusion on how to exponentially speed up spectroscopic mapping could then be phrased as: "measuring fewer spectra faster".

We exemplify the implementation of this concept using topographic mapping in which a full spectrum would have traditionally been recorded at every topographic location. This measurement mode is referred to as quasiparticle interference imaging(*8–16*) and it provides insight into the band-structure of quantum materials from a Fourier transform of local density of states (LDOS) maps. QPI imaging requires the recording of a massive amount of spectra that can occupy an STM for several days for a single QPI dataset(*13*). In the spirit of the trivial conclusion, we utilize our sparse sampling approach(*17*) and combine it with a parallel spectroscopy(*18*) addition to the STM to exponentially enhance spectroscopic mapping. While the former enables the recording of QPI maps using fundamentally fewer LDOS measurements, the latter works by increasing the speed of an individual spectrum.

Figure 1a illustrates our setup, which was reversibly added to an existing low-temperature STM (Createc) using a relay switch, a compensating capacity, and a multifrequency lock-in amplifier (Intermodulation Products, MLA-3)(SI). The switch(*19*) enables to toggle between conventional STM operation and parallel spectroscopy for which we require the compensating capacity ($C_c$) to remove stray capacity related displacement currents using $V_{cmp}$ before reaching the preamplifier(*18*)(SI). The sparse-sampling concept requires no additional hardware but only the control of the tip motion on a subset of randomized locations along a near optimal path (purple line) that was created in the spirit of a traveling salesperson(*17*). Note that for other spectroscopic mapping implementations, the locations would be substituted with temperature, magnetic field, doping, etc.

The subsampling rate directly relates to the total mapping time and offers potential for further speed enhancements compared to conventional mapping when sparsity is high and sample symmetries are fully exploited(*17*). At this point, it is crucial to realize that STM investigations are amenable to compressive sensing concepts(*20, 21*) because the data structure is oftentimes highly sparse or compressible in one representation space. In our example, this applies to the QPI pattern(*17, 22*), which consists of only a few nonzero values in Fourier domain (Fig. 1b).

To speed up the measurement at each of those reduced number of locations, a harmonic drive $V_{drv}$ is applied onto the tunneling junction which is characterized by nonlinearities in the current-voltage (*I-V*) characteristic (Fig 1c). These nonlinearities lead to the generation of higher order harmonics (Fig 1d) whose parallel measurement (here 31) enables the precise reconstruction of their generating *I-V* signature via an inverse Fourier transform(*18*). Since the drive amplitude covers the desired energy range, this concept yields the full spectrum in a fraction of the time when compared to conventional spectroscopy that consisted in incrementally sweeping a voltage and recording the corresponding current/conductance. As QPI mapping requires LDOS measurements, which are proportional to the conductance, we also show a comparison of the conventional (blue) and parallel measurement (orange dashed) of the latter (Fig 1e), proving their excellent theoretical agreement.

Figure 2 exemplifies how either of the two trivial speed enhancements work independently for QPI imaging. In this example, we map the Shockley surface state of Au(111) by only using parallel spectroscopy (top row) or only sparse sampling (bottom row) on grids that previously could have been considered impossible. In the top-row, we move the tip along a regular grid but measure each pixel using parallel spectroscopy. The speed enhancement solely comes from the faster point spectrum, taking only about 100 ms. On the bottom-row, we use sparse-sampling on a small selection of randomly chosen points but measure every spectrum using conventional point spectroscopy. Here, speed enhancement comes from the reduced number of measurements. In both modes, we clearly see the standing wave patterns (a, d), showing the scattering of electron waves at discontinuities and which lead to the ring in the QPI maps (b, e). The energy resolved profiles (c, e), show the near-free electron like behavior of the charge carriers in Au(111) and nondispersing features that are related to the $(22 \times \sqrt{3})$ herringbone reconstruction(*23*). The latter acts as a periodic potential responsible for the opening of an energy gap around -312 mV(*24*, *25*) (purple arrow in e). Our measurements not only prove the independent viability of parallel spectroscopy and sparse sampling but they also reveal sufficient detail to access the physics of band-folding that is difficult to spot with ARPES(*26*) and that has received considerable attention by the advent of twisted bilayer systems(*27*). This further reinforces the singular role of QPI mapping for the high-resolution exploration of quantum materials.

In Figure 3, we conclude our trivial quest and combine parallel spectroscopy with sparse sampling. In addition to Au(111) (a-c), we demonstrate rapid QPI mapping with Bi2212 as one of the cuprate's most notorious examples(*28*, *8*, *29*, *10*, *11*, *30*, *12*, *16*, *31*, *32*). Our combination of sparse sampling and parallel spectroscopy enables the compromise-free spectroscopic mapping of a large reciprocal space that we thoroughly map within only a few hours in what would have been previously a monthlong measurement. The QPI maps of Bi2212 reveal the Bragg peaks of the CuO lattice (e) and the dispersion plots (f) capture the dispersion relation along the indicated reciprocal directions. The presence of Bragg peaks shows the excellent compatibility of our method with established data processing(*12*) and analysis approaches.

The potential of a $10^4$ speed advantage can be unlocked by further enhancing the sparse sampling and parallel spectroscopy contributions as the slightest improvement of either will have immediate impact on the total measurement time through their geometric relationship. For the sparse sampling case, the exploitation of sample symmetries could lead to hundred-fold enhancement(*17*); while parallel spectroscopy becomes faster, the faster $V_{drv}$ and the more demodulators are used(*18*).

Finally, our concepts can be easily extended to other spectroscopic modes in which the location would be replaced by charge-carrier concentration or magnetic field. Therein, one would measure the spectroscopy at a random subset of that variable and utilize a suitable sparsifying basis for compressive sensing. Instead of the Fourier space, this could be a wavelet-based domain or a dictionary that properly encodes the intrinsic structure of the measured data, making also these mapping modes amendable to substantial speed enhancements.

Since experimental discovery is ideally a process in which first impressions and trends of early measurements inform further steps in the exploration, our faster mapping concept will bring active control at the hands of the experimenter. This will not only provide a faster convergence to the understanding of novel materials, but is also immediately practical to obtain proper statistics and coping with experimental challenges, such as tip-related instabilities, noisy lab settings, or the limited hold-time of ones cryostat.

**Acknowledgments:** We acknowledge fruitful discussions with Riccardo Borgani and Daniel Forchheimer. We thank Qisi Wang and Johan Chang for their help with mounting the Bi2212 sample. **Funding:** F.D.N. thanks the Swiss National Science Foundation under Project Number PP00P2-176866 and ONR (N00014-20-1-2352) for generous support. D.L. thanks UZH Forschungskredit (FK-20-093), and T.K. acknowledges support by JSPS KAKENHI (Grants No. 19K03733). **Author contributions:** F.D.N. conceived and supervised the work. B.Z. implemented parallel spectroscopy and J.O. sparse sampling. F.D.N., D.L., and B.Z. carried out the measurements. T.K. synthesized the Bi2212 crystal. D.L. and B.Z. prepared the samples. F.D.N., B.Z., J.O., and L.N. analyzed the data. F.D.N. wrote the manuscript. **Competing interests:** Authors declare no competing interests. **Data and materials availability:** All data is available in the main text or the supplementary materials.

**Supplementary Materials:**

Materials and Methods:

<u>Sample and tip preparation</u>

We obtain atomically clean Au(111) surfaces after repeated cycles of sputtering (Ar$^+$ ions, 4 µA/cm$^2$, 20 minutes) and annealing (850 K for 20 minutes). We perform our QPI mapping on terraces whose widths exceed 200 nm at 77 and 4.2 K. The point-impurities at which the surface state is scattered are of unknown origin.

The Bi$_2$Sr$_2$CaCu$_2$O$_{8+\delta}$ sample (Bi2212 at a doping level of 0.14, 1.5 mm by 1.5 mm) was glued together with a cleaving post onto a sample holder, transferred without any further heat-treatment into our ultra-high vacuum system, and cleaved *in-situ* prior to transferring into the cooled scanning tunneling microscope.

We used an electrochemically etched tungsten tip that we treated by e-beam heating, sputtering and poking into Au(111), until it showed a robust metallic *I-V* characteristic.

<u>Displacement Current Compensation:</u>

Since the parallel spectroscopy method involves the excitation of the tunnel junction with a large voltage amplitude $V_{\text{drv}}(t) = V_{\text{drv}} cos(f_{\text{drv}} t)$, any stray capacitance $C_s$ will lead to displacement currents via $I_{\text{dsp}} = C_s \dot{V}_{\text{drv}}$ that may overwhelm the current preamplifier. We therefore actively cancel $I_{\text{dsp}}$ (*18*) with a compensating capacitor $C_c$ that we place in-parallel to the input of the current-preamplifier (Femto, DLPCA-200, 50 kHz setting) and to which we apply a compensating voltage $V_{\text{cmp}}(t) = -V_{\text{drv}}(t)\, C_c/C_s$, see Figure 1a. Our multifrequency lock-in amplifier (Intermodulation Products, MLA-3) provides two phase synchronized outputs which we use for $V_{\text{drv}}$ and $V_{\text{cmp}}$ (Figure 1a). We find the nearly ideal phase relationship φ for proper cancellation, close to φ = π, and that we further refine by an iterative procedure which adjusts $V_{\text{cmp}}$ as well as φ. In view of the tip-height dependence of the stray capacitance, we calibrate our compensation just outside the tunneling range by lifting the tip off the surface by about 6-10 nm, which is a range where the tip-related capacitance changes are negligible(*33*).

Sparse sampling measurement matrices:

Our sparse sampling concept(*17*) uses a random path of topographic locations where we record the LDOS via conventional or parallel spectroscopy. In order to create an efficient routing between those locations, we use a genetic traveling salespersons (TSP) implementation(*34*) to find a near optimal solution to reduce measurement overhead related to tip-motion. In the present work, we create paths having between 32'000 and up to 125'000 locations. The calculation of such a large path problem is computationally too expensive for us but it can be simplified using the concept we had suggested earlier(*17*). We divide the locations into several smaller sectors for which we independently and rapidly find a TSP solution. We then combine these sectors using a second TSP solution, allowing us to calculate paths for a large number of locations with ease. Note that all of our paths were created with fewer locations in the vicinity of the map-edges to mitigate Fourier artifacts, as described before(*17*).

We recover the QPI information from the LDOS measurements and the knowledge of the topographic locations(*17*) using the compressive sensing solver SPGL1(*35*).

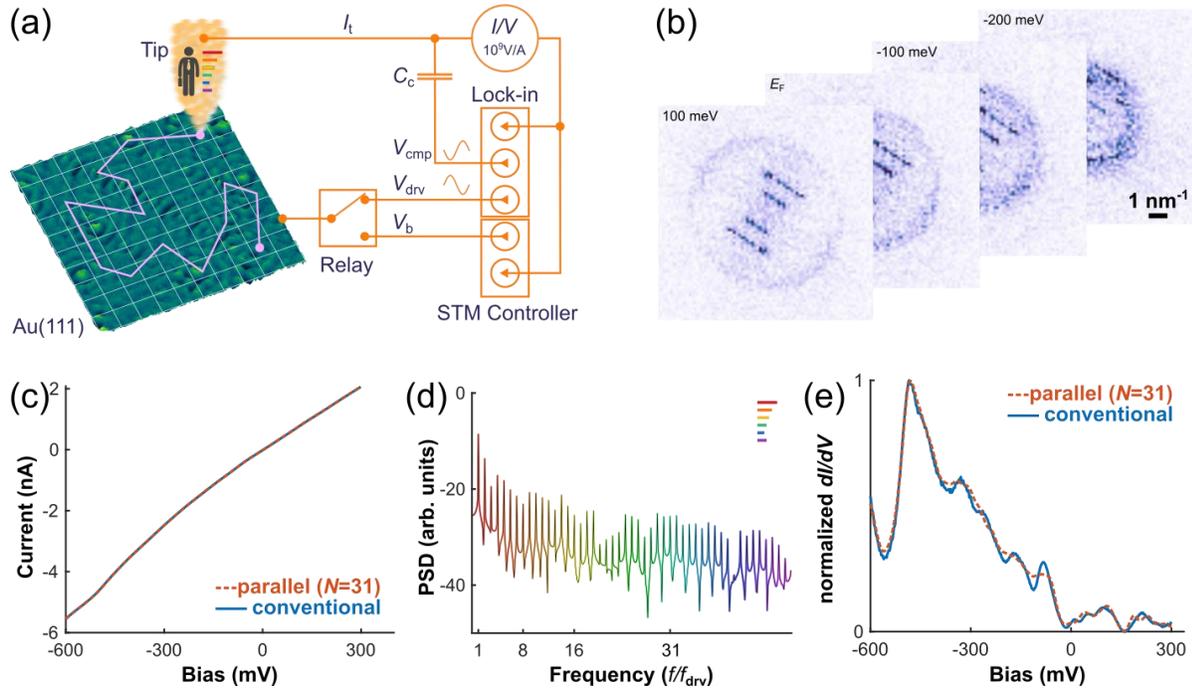

**Fig. 1. Experimental setup and working principles used for fast spectroscopic mapping. (a)** Wiring schematics of the components added to an existing STM setup, including a multifrequency lock-in amplifier, a relay switch, and a compensating capacitor. The purple line symbolizes the traveling salesperson's path used for sparse sampling. **(b)** Simultaneously obtained quasiparticle interference patterns from Fourier transforms of LDOS maps, showing the Au(111) surface state and indicating the sparsity in the QPI domain (setpoint: $V_b$ = -200 mV, $V_{drv}$ = 600 mV, $f_{drv}$ = 250 Hz, $I_t$ = 800 pA, $T$ = 77 K). **(c)** Current-voltage characteristics (*I-V*) of Au(111) measured with conventional spectroscopy (blue). The *I-V* curve is characterized by one dominant and a few smaller nonlinearities that, upon application of a harmonic drive, generate **(d)** higher order harmonics, shown here by their power-spectral density. Their simultaneous demodulation (here 31 harmonics are used) enables the reconstruction of the original *I-V* characteristics (orange dashed line in **c**) and **(e)** the conductance (orange dashes), shown here in comparison with the conventionally measured spectrum (blue).

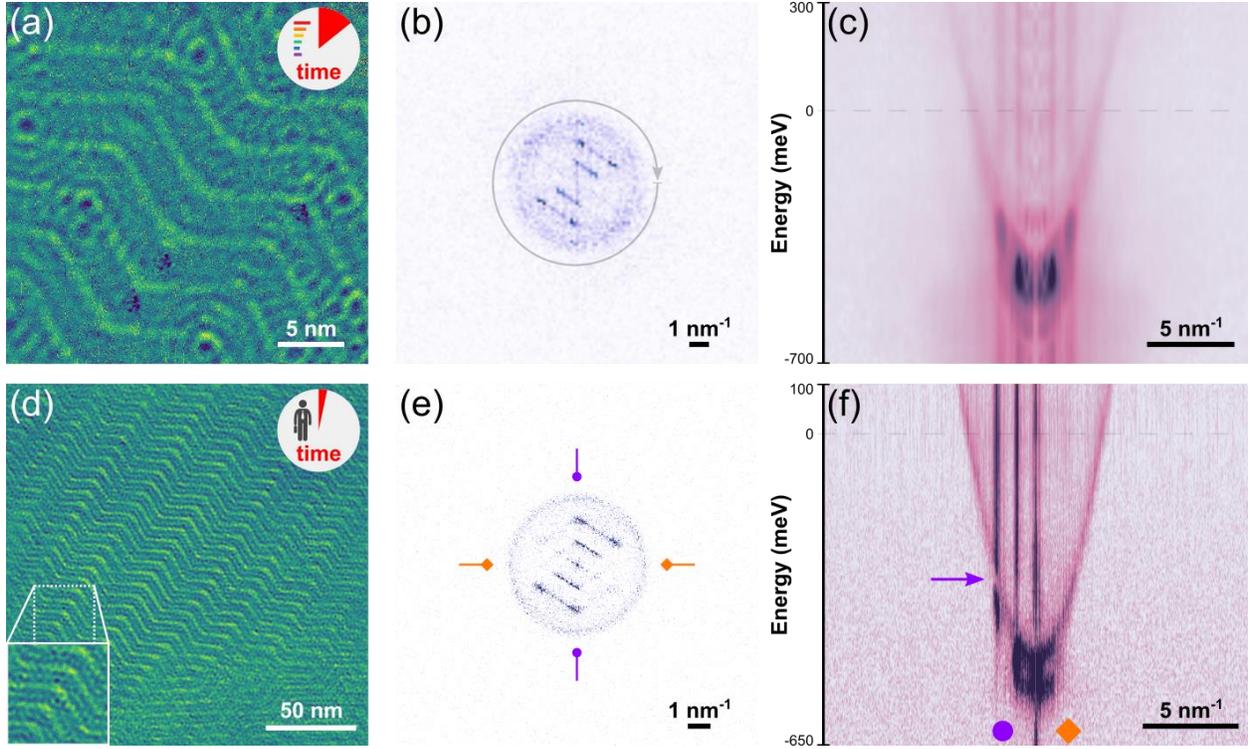

**Fig. 2. Parallel Spectroscopy and Sparse Sampling for quasiparticle interference imaging, independently exemplified with Au(111). (a)** LDOS at -100 mV, measured using parallel spectroscopy on a regular 512 × 512 grid, taking 10.3 hours instead of 73 hours (setpoint: $V_b$ = -200 mV, $V_{drv}$ = 600 mV, $f_{drv}$ = 250 Hz, $I_t$ = 800 pA, $T$ = 77 K). **(b)** QPI pattern obtained from a Fourier transform of **a**. **(c)** Dispersion relation from radially averaged QPI maps showing the nearly-free electron band ($m^*/m_e$ = 0.24 ± 0.02) and the nondispersing herringbone reconstruction. **(d)** LDOS at -100 mV, obtained using a 2% subsampled 2048 × 2048 grid, taking about 35 hours, instead of 48.5 days (setpoint: $V_b$ =-700 mV, $I_t$ = 6 nA, $V_{mod}$ = 5 mV, $f_{mod}$ = 997 Hz, $T$ = 4.2 K). **(e)** QPI pattern obtained from our sparse reconstruction. **(f)** Dispersion relation along the directions indicated in **e** ($m^*/m_e$ = 0.221 ± 0.002). Note the presence of a bandgap when the surface state coincides with the herringbone periodicity around -312 mV (arrow), reflecting the back-folding of the surface state due to a periodic potential.

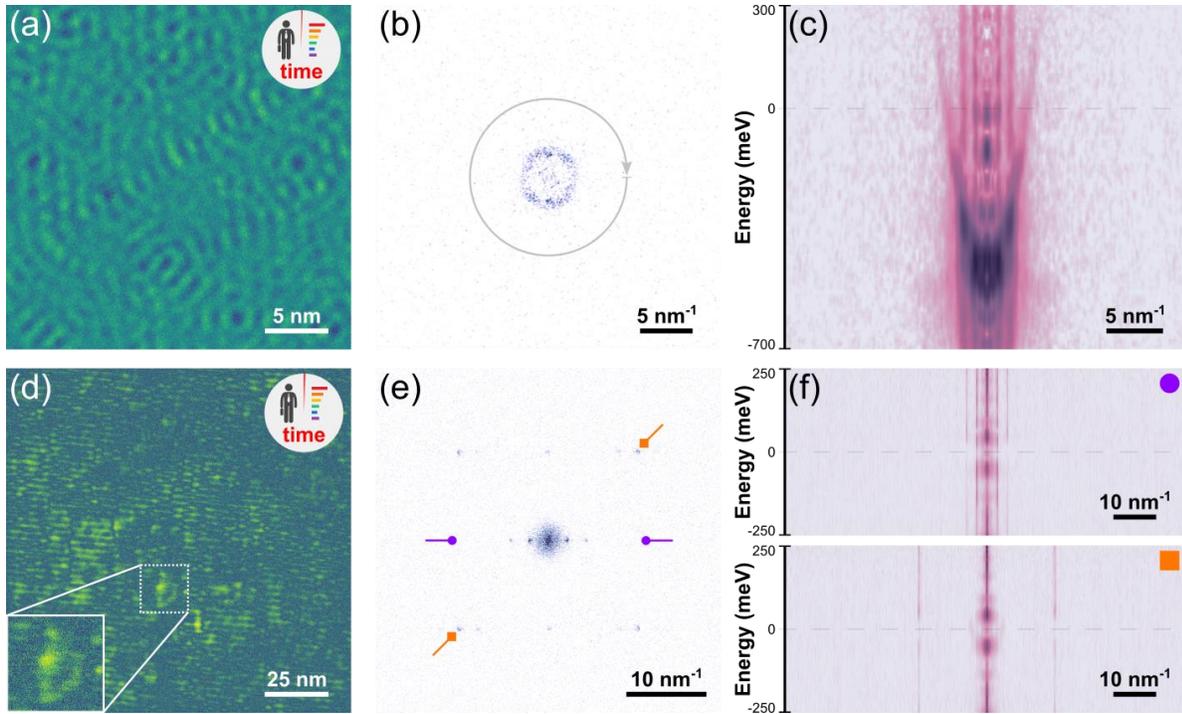

**Fig. 3. Combination of Parallel Spectroscopy and Sparse Sampling (a)** LDOS at -180 mV, measured using the combined concept of parallel spectroscopy and sparse sampling on a 3% subsampled 1024 × 1024 grid on Au(111), taking 73 minutes instead of 12 days (setpoint: $V_b$ = -200 mV, $V_{drv}$ = 600 mV, $f_{drv}$ = 250 Hz, $I_t$ = 800 pA, $T$ = 77 K). **(b)** QPI pattern obtained from **a**. **(c)** Dispersion relation from radially averaged QPI maps. **(d)** LDOS of Bi2212 at 100 mV, as obtained from the combination of parallel spectroscopy and sparse sampling on a 3% sampled 2048 × 2048 grid, taking 6.5 hours, instead of 48.5 days. The inset shows atomic contrast, also discernible by the Bragg peaks in **e** and **f** (setpoint: $V_b$ = -100 mV, $V_{drv}$ = 400 mV, $f_{drv}$ = 250 Hz, $I_t$ = 400 pA, $T$ = 4.2 K). **(e)** Raw QPI pattern of Bi2212 at 100 mV, as obtained after sparse reconstruction of the parallel spectroscopy enhanced measurement. **(f)** Dispersion relation along the lines indicated in **e**. The time comparisons assume 1 second per spectrum for conventional mapping.